\documentclass{article}

\usepackage{arxiv}

\usepackage{cite}
\usepackage{amsmath,amssymb,amsfonts}
\usepackage{algorithmic}
\usepackage{graphicx}
\usepackage{textcomp}
\usepackage{epstopdf}
\usepackage[T1]{fontenc}
\usepackage{lmodern}

\usepackage{csvsimple}

\usepackage{verbatim} %for block commenting
\usepackage{soul} %for text highlighting
\usepackage{wasysym} %for smileys
\usepackage[T1]{fontenc}
\usepackage[utf8]{inputenc}
\usepackage[emoticons]{emoji}

\usepackage{graphics}

\title{The Potential of Machine Learning and NLP for Handling Students' Feedback – A Short Survey}

\author{
 Maryam Edalati\\
 Dept. of Computer Science, Norwegian University of Science \& Technology (NTNU), Norway.\\ maryame@stud.ntnu.no\\}

\begin{document}
\maketitle

\begin{abstract}
This article provides review of the literature of students' feedback papers published in recent years employing data mining techniques. In particular, the focus is to highlight those papers which are using either machine learning or deep learning approaches. Student feedback assessment is a hot topic which has attracted a lot of attention in recent times. The importance has increased manyfold due to the recent pandemic outbreak which pushed many colleges and universities to shift teaching from on-campus physical classes to online via eLearning platforms and tools including massive open online courses (MOOCs). Assessing student feedback is even more important now. This short survey paper therefore highlights recent trends in the natural language processing domain on the topic of automatic student feedback assessment. It presents techniques commonly utilized in this domain and discusses some future research directions. 

\end{abstract}


%%%%%%%%%%%%%%%%%%%%%%%%%%%%%%%%%%%%%%%%%%%%%%%%%
%% INTRODUCTION
%%%%%%%%%%%%%%%%%%%%%%%%%%%%%%%%%%%%%%%%%%%%%%%%%

\section{Introduction}
\label{sec:introduction}

The main objective of this article is to review the literature concerning the use of Natural language processing (NLP) in student feedback. Feedback analysis is crucial for higher education since it could operate effectively on the curriculum. Teaching quality would significantly improve if the professors have a clear idea about how well different aspects of teaching such as content, structure, assignment and teaching methods, etc., work.

Traditionally in the universities and colleges, students could give feedback on the university website or in some cases on the printed forms. However, the growth of online course platforms such as MOOC and the issues related to it \cite{dalipi2016towards}, for instance, higher dropout rates \cite{dalipi2018mooc,imran2019predicting} increased the demand for incorporating students' feedback dramatically. Many higher education institutes and experts have had a strong interest to extract aspects and their related sentiment from this feedbacks \cite{Do2019,Kastrati2020}. Manual aspects and its related sentiment extraction is time-consuming due to a large number of data. Therefore developing a reliable automated method to extract aspects and related sentiment of the aspect is necessary \cite{Sindhu2019}.

Opinion mining is an effective method that is able to overcome the limitation of the old methods to exploit information that expressed through the feedbacks. Opinion mining (OM) or Sentiment Analysis (SA), extracts the user's opinions (sentiment) from a target text and pointed out their related polarity. In recent years, sentiment analysis have been applied to a variety of tasks including examining the spreading pattern of information and tracking/understating public reaction during given crisis on social media \cite{Imran2020,Kaila:2020}.

\begin{figure}[ht] 
  \centering
  \includegraphics[width=\textwidth]{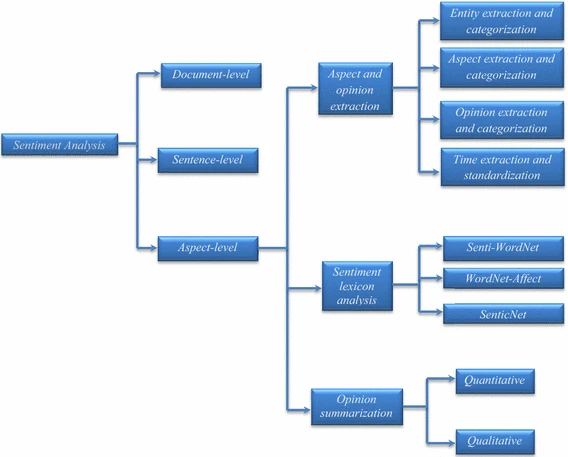}
  \caption{A taxonomic view on sentiment analysis techniques~\cite{Rana2016}.}  
  \label{fig:SA_classification}
\end{figure}

SA could be studied at three different levels: document-level, sentence-level, and entity or aspect-level~\cite{Hu2014,Liu2012}.
The assumption in document and sentence level SA is on that only one topic expressed in the document or sentence. However, in many situations, this is not the case and a precise analysis requires also investigation at the entity and aspect level~\cite{Do2019}. In aspect-level SA, first the aspects and opinions are extracted then classified them into similar classes, second the polarity of the opinions determined and summarized the results. Figure~\ref{fig:SA_classification} demonstrated the classification levels and detailed about aspect-level SA~\cite{Rana2016}.

%%%%fig:SA_classification
%%%%ùA taxonomic view on sentiment analysis techniques~\cite{Rana2016}.

Although in the recent decade a huge number of study conducted in  different classification levels, we narrow down this survey on the aspect level classification between years 2019-2020.

Hemmatian et al.~\cite{Hemmatian2019} have done a review on opinion mining area and its related classification techniques. Rana et al.~\cite{Rana2016} surveyed different techniques for the extraction of aspects from online reviews.

One of the forms of distance learning is e-learning. E-learning has become popular in recent years because of development in the technology \cite{imran2012multimedia, imran2014hip} and the utilization of NLP techniques to create effective learning management systems and elearning platforms \cite{imran2012semantic}. In 2020 due to the COVID19 and its special circumstances, many educational institutes worldwide move completely their operations to e-learning. Massive Open Online Courses known as MOOCs are one of the first e-learning platforms. MOOCS are open-access online courses that allow for unlimited participation, as well as Small Private Online Courses (SPOCs)~\cite{Kaplan2016}. MOOCs offer a great platform to collect students feedback on a massive scale and to train and build models on.

The rest of the article is as follows. In Section~\ref{sec:relatedWork}, the most recent works is presented. Section~\ref{sec:classification} presented the techniques and approaches which have been used to conduct sentiment analysis and Aspect Based Sentiment Analysis (ABSA). Following that in Section~\ref{sec:DR} and section~\ref{sec:rd}, Dataset Resources and Research Directions discussed respectively. Finally in the Section ~\ref{sec:conclusion} the conclusion and future directions are presented. 

\section{Literature Review}
\label{sec:relatedWork}
 Kastrati et al.~\cite{Kastrati2020AspectBasedOM} proposed a model for aspect-based opinion mining. They collect the dataset containing more than 21 thousand reviews on the Coursera. All reviews were in English. Authors used three different techniques as representation techniques: Term Frequency (TF), and term frequency-inverse document frequency (tf*idf), and word embedding. They first classified the comments based on five aspect categories including Instructor, Content, Structure, Design and General, then the aspects classified based on the polarity [Positive, Negative, and Neutral]. Four conventional machine learning classifiers, namely Decision Tree, Naïve Bayes, SVM, and Boosting and a 1D-CNN model were used. The result shows conventional machine learning techniques achieved better performance than 1D-CNN.
 
Sangeetha et al.~\cite{Sangeetha2020} used the Vietnamese students feedback corpus (UIT-VSFC) dataset. The dataset is in Vietnamese that converted to English language for their work. The dataset consisting of 16,175 students feedback sentences. They focus to sentiment classification [positive, negative and neutral]. In their proposed method, input sequences of sentences are processed parallel across multi-head attention layer with fine grained embeddings (Glove and Cove) and tested with different dropout rates to increase the accuracy then the information from both deep multi-layers is fused and fed as input to the LSTM layer. They compered their proposed method with the other baseline models [LSTM, LSTM + ATT, Multi-head attention]. The result shows the proposed method shows better over all result. 

Nikoli et al.~\cite{Nikoli2020} proposed a method for the Aspect-based sentiment analysis in the Serbian language at the sentence segment level. They used a dataset that contain both official Faculty and online surveys. The dataset labeled for 7 aspect classes [professor, course, lectures, helpfulness, materials, organization and other] and 2 polarity classes [positive, negative]. Authors used term frequency-inverse document frequency (tf*idf) as representation techniques. For classification they used three standard Machine learning  multi-class classification model (Support vector machine, k-nearest neighbours (k-NN), and multi-nomial NB (MNB), and a cascade classifier including set of SVM classifiers organized in a cascade structure. The results indicates F score for the aspects classification vary between 0.49 and 0.89. The F score is 0.83 for positive sentiment and 0.94\% for negative sentiment. 

Sindhu et al.~\cite{Sindhu2019} proposed a supervised aspect based opinion mining system based on two-layered LSTM model so that the first layer predicts the six categories of aspects [Teaching Pedagogy, Behavior, Knowledge, Assessment, Experience, and General] and second layer predict polarity [positive, negative, and neutral] of aspect. The authors used a data set of last five years students feedback of Sukkur IBA University for this study. The accuracy for the aspect extraction and sentiment polarity detection are 91\% and 93\% respectively. 

Kastrati et al.~\cite{Kastrati2020} proposed an aspect sentiment analysis to automatically identify sentiment or opinion polarity expressed towards a given aspect related to the MOOC. Their methods takes advantage of the weak supervision strategy to train a deep network to automatically identify the aspects on MOOC by using either very few or even no manual annotations. In addition, the proposed framework examines the sentiment towards the aspects commented on a given review. The result shows F1 score of 86.13\% and 82.10\% for aspect category identification and aspect sentiment classification respectively.

Lundqvist et al.~\cite{Lundqvist2020} have conducted a research to evaluate student's feedback within a MOOC. They worked on the dataset containing 25,000 reviews from MOOC's users. The participants divided into three groups (beginner, experienced, and unknown) based on their level of prior experience. The VADER (Valence Aware Dictionary for sEntiment Reasoning) sentiment algorithm was used for sentiment analysis.

Shaikh et al.~\cite{Shaikh2019} proposed a two-step strategy based on Machine Learning and Natural Language Processing (NLP) techniques to extract the aspect and polarity of the feedback text respectively. They used 10,000 labeled student feedbacks collected at Sukkur IBA University Pakistan. Their methods is divided into three main steps. In the first step, the student feedbacks are classified into the teacher or course entity by using the Naive Bayes Multinomial classifier. Once the entity has been extracted, a rule-based system developed to analyze and extract aspects as well as opinion words from the text by using predefined rules. In the final step, the authors used SentiWordNet to extract the sentiment regarding extracted aspects. The result shows first, the overall precision of 83.89\% and 84\% on the teacher and course entity respectively. Second the overall precision of 83\% and recall of 80\% for extracting different aspects of teacher and course and finally 90\% accuracy for the sentiment classification. 

Authors in~\cite{MARCU2020} conducted a document-level sentiment analysis. They used a dataset containing 191 students' feedback. The data analyzed through the Orange environment (an open source tool for machine learning and data visualization). They made a comparative study of the obtained results using the Ekman and Plutchik models. Five emotions anger, disgust, fear, joy, sadness and surprise have been studied. The result shows almost 37\% of the documents were marked identically by the two models.
Table~\ref{tab:ReviewStudies} lists some of the reviewed papers detailing the level, review representation, classification model, and the Language.

\begin{figure}[ht] 
  \centering
  \includegraphics[width=\textwidth]{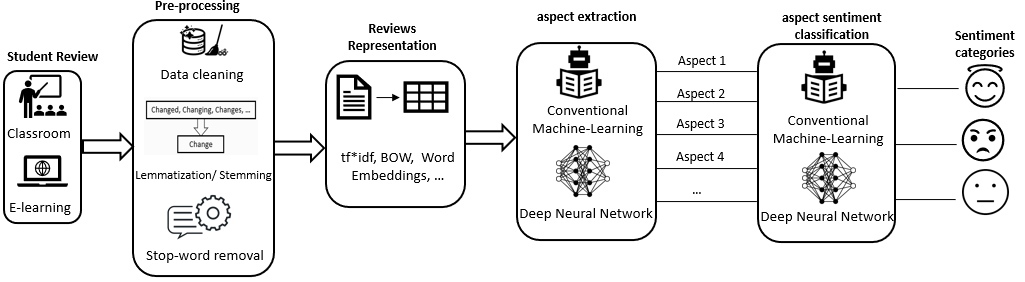}
  \caption{A schematic of the aspect-level sentiment analysis.}  
  \label{fig:schematic}
\end{figure}

\section{Representation \& Classification Approaches}
\label{sec:classification}
Despite a variety of models that have been used in different researches all the papers used the same architecture for extraction and classification of the aspect. Figure~\ref{fig:schematic} illustrated the different steps in aspect-based SA. Aspect-based SA generally consists of four main steps: Pre-processing, Review Representation, Aspect Extraction, and Aspect Sentiment classification. We will discuss all these steps separately in the following subsections. 

\subsection{Pre-processing}
After collecting the students' feedback from either the online platforms or traditional classrooms, the raw data included html tags (for reviews
were scraped from the web)~\cite{Kastrati2020AspectBasedOM}, spelling mistakes, emoticons, icons, and unwanted symbols like~(!,@,\&,\#,\}), repeated letters, numbers and stop words~\cite{Sangeetha2020}. In order to prepare the raw data for the feature extraction step, these unwanted elements need to removed from the data and lower-casing the sentences. In the end, the text normalization (reducing words to their word root part) is done by either stemming~\cite{Kastrati2020AspectBasedOM,SRINIVAS:2019,estrada2020opinion} and lemmatization. Nikolić et al.~\cite{Nikoli2020} have done the preprocessing by removing stop words (the most common words in the Serbian language), stemming and various n-gram (contiguous sequences of n words) lengths and frequencies.  
  
\subsection{Representation of Reviews}
Since machine learning/deep learning algorithm couldn't feed by the text, the text should convert  to the numerical format (vector). There are many algorithms that can be used to convert text data to vector of numbers: 
Bag-of-Words (BoW)~\cite{Kastrati2020AspectBasedOM,Kandhro:2020}, term frequency(tf)~\cite{Kastrati2020AspectBasedOM,Lwin:2020}, term frequency inverse document frequency(tf*idf)~\cite{Kastrati2020AspectBasedOM,Nikoli2020,Moharil:2020,SRINIVAS:2019,Hariyani:2019}, Word embedding. Semantics, on the other hand, can play a vital role in identifying correct aspects employing objective metric~\cite{kastrati2015semcon}. These are discussed in next section.

\subsection{Semantic Techniques}

Semantics has been widely used in the education domain, for instance, a document level review can be enriched with semantics~\cite{kastrati2015general} using background knowledge provided by an ontology~\cite{kastrati2015using} and through the acquisition of its relevant terminology~\cite{kastrati2019impact}. Further objective metrics~\cite{kastrati2016semcon} could be exploited for sentiment analysis. It is used to enrich existing concepts in domain ontologies for describing and organizing multimedia documents~\cite{kastrati2019integrating}. Semantics can be represented as word embeddings, that are one of the key breakthroughs in the NLP. Word Embeddings are d‐dimensional space representations of words' distributed. As represented in Figure~\ref{fig:schematic} word embeddings are the first step in DL architecture after the pre-processing step. The output of the word embedding step used as input for the DL architecture for either extracting the aspect or classification of sentiment. 
Following pre-trained word embedding Word2Vec~\cite{mikolov2013efficient} which have developed by Google in 2013, other word embedding methods such as GloVe~\cite{Pennington:2014} and fastText~\cite{bojanowski:2017} developed by Stanford University and Facebook respectively. In 2018 Devlin et al.~\cite{Devlin:2018} developed Bidirectional Encoder Representations from Transformers (BERT). 

Word2Vec ~\cite{Kastrati2020AspectBasedOM, Sindhu2019,Kastrati2020}, GloVe~\cite{Kastrati2020AspectBasedOM,Sangeetha2020,Kastrati2020,Mostafa:2020}, 
FastText~\cite{Kastrati2020AspectBasedOM,Kastrati2020} are the most used word embedding in student feedback SA. Estrada et al.~\cite{estrada2020opinion } and Sangeetha et al.~\cite{Sangeetha2020 } used BERT and Contextualized word vector for embedding respectively.

\begin{figure}[ht] 
  \centering 
  \includegraphics[width=\textwidth]{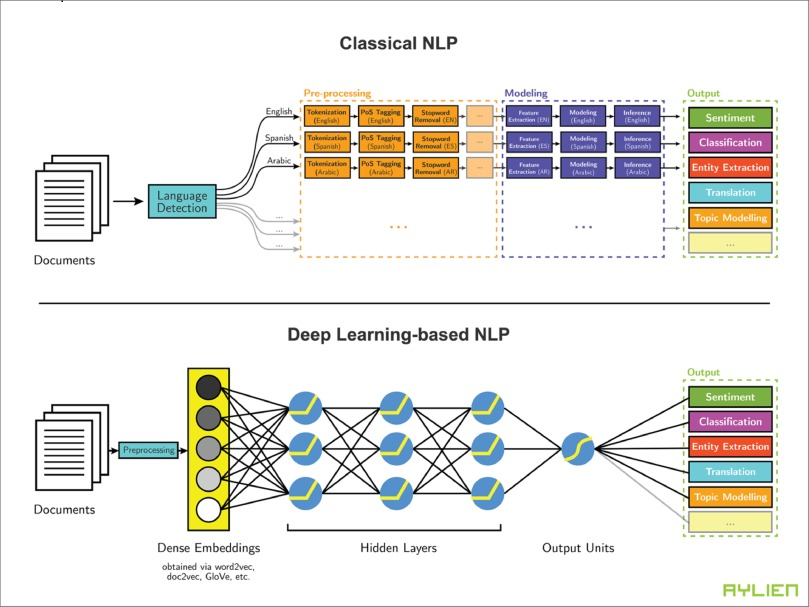}
  \caption{A comparison between conventional machine learning techniques and deep learning for natural language~\cite{Do2019}.}  
  \label{fig:schematic}
\end{figure}

\subsection{Conventional ML Techniques}
Generally, the classification techniques for SA divide into two main categories: 
\begin{itemize}
  \item Supervised methods. In this method labels according to the corresponding characteristics are available and the model would be learned by using the labels. 
  \item Unsupervised methods. Unsupervised methods try to classify data by discovering patterns in unlabeled data.
\end{itemize}
When it comes to SA, classification methods divided into two main groups: Conventional ML methods and deep learning methods. 
Figure~\ref{fig:schematic} illustrates the comparison between the conventional ML techniques' steps and deep learning techniques' steps.
Finding an effective method for aspect extraction and aspect polarities classification is a challenging task. Many pieces of research conducted on the comparison between different methods. In comparison between the conventional machine learning methods and deep learning methods, conventional methods such as Decision Tree~\cite{Kastrati2020AspectBasedOM}, Logistic Regression~\cite{Lwin:2020} and evolutionary approach called EvoMSA~\cite{estrada2020opinion} show better results.
Conventional ML methods such as Support vector machine (SVM)~\cite{Kastrati2020AspectBasedOM,Nikoli2020,Moharil:2020,estrada2020opinion,Lwin:2020,SRINIVAS:2019,Hariyani:2019}, Naïve Bayes (NB)~\cite{Kastrati2020AspectBasedOM,Nikoli2020,Shaikh2019,estrada2020opinion,Hariyani:2019}, Random Forest (RF)~\cite{estrada2020opinion,Lwin:2020}, Decision Tree (DT)~\cite{Kastrati2020AspectBasedOM,Hariyani:2019} and, K-Nearest Neighbor (KNN)~\cite{Nikoli2020,Hariyani:2019} are the most popular methods for SA.
Other conventional machine learning methods such as Boosting~\cite{Kastrati2020AspectBasedOM}, Bernoulli Naive bayes~\cite{estrada2020opinion}, K-means~\cite{SRINIVAS:2019}, logistic model tree (LMT)~\cite{Lwin:2020} more or less used in the different researches.

\subsection{Deep Learning Techniques}
Deep Learning Techniques used widely in SA because they are nonlinear and  could  process a large amount of data  natural data in their raw form~\cite{LeCun:2015}. Deep learning is a machine learning method that allows computational models with multiple processing layers to learn representations of data with multiple levels of abstraction~\cite{LeCun:2015}.
As illustrated in Figure~\ref{fig:schematic}, Deep Neural Network (DNN) methods for SA consist of three parts: dense word embedding, multiple hidden layers between the input and output and output units~\cite{Do2019}.

The limitation of the other deep learning architectures such Recurrent Neural Networks (RNN) has been improved with the introduction of networks such as long short-term memory (LSTM)~\cite{Do2019}. The authors in~\cite{Sindhu2019,Kastrati2020,estrada2020opinion,Kandhro:2020,Lwin:2020,Mostafa:2020,Nimala:2019 } used LSTM architecture. The basis of LSTM is a memory cell that controls the read, write and reset operations of its internal state through output, input and forget gates~\cite{Do2019}.
Convolution neural networks (CNN) are the other most used architecture~\cite{Kastrati2020AspectBasedOM,Kastrati2020,estrada2020opinion}. Since CNN is a a non-linear supervised model, it could better fit the data compare with the linear model and it also does not need extensive hand-crafted features such as fixed language rules~\cite{Poria:2016}. Poria et al.~\cite{Poria:2016} also showed that deep CNN is more efficient for aspect extraction than existing approaches.

\begin{table}[ht]
\centering
\begin{tabular}{|p{1.5cm}|p{0.8cm}|p{1.5cm}|p{2.5cm}|p{5.5cm}|p{1.6cm}|}
\cline{1-6}
Author             & Year &  Level           & Review Representation & Classification Models & Language   \\ \cline{1-6}
Kastrati et al.{\cite{Kastrati2020AspectBasedOM}}  & 2020    & aspect    & Word embeddings, BOW  & DT, NB, SVM, Boosting , CNN & English    \\ \cline{1-6}
Sangeetha et al. {\cite{Sangeetha2020}} & 2020  & sentence & Word embeddings       & Fusion multi-head attention                                             & Vietnamese \\ \cline{1-6}
Nikolić et al. {\cite{Nikoli2020}}    & 2020& aspect    & tf*idf     & k-NN, NB, SVM & Serbian    \\ \cline{1-6}
Sindhu et al. {\cite{Sindhu2019}}    & 2019  & aspect    & Word embeddings       & Two-layered LSTM                                                        & English    \\ \cline{1-6}
Kastrati et al. {\cite{Kastrati2020}}  & 2020  &aspect   & Word embeddings  & CNN and LSTM                                                            & English    \\ \cline{1-6}
Shaikh et al. {\cite{Shaikh2019}}    & 2019 & sentence  & String2WordVector     & Naive Bayes Multinomial classifier & English    \\ \cline{1-6}
Marcu et al. {\cite{MARCU2020}}     & 2020 & document  & Orange environment    & Ekman and Plutchik  & Romanian   \\ \cline{1-6}

Moharil et al. {\cite{Moharil:2020}}     & 2020 & document  & tf*idf    & SVC-linnear classifier  & English   \\ \cline{1-6}
Estrada et al. {\cite{estrada2020opinion}}     & 2020 & sentence  & BERT    & SVC-linnear classifier, SVM; Bernoulli Naive bayes, Random Forest, LSTM, CNN  & English   \\ \cline{1-6}
Kandhro et al. {\cite{Kandhro:2020}}     & 2020 & document  & BoW, Word2Vec    & LSTM  & English   \\ \cline{1-6}

Lwin et al. {\cite{Lwin:2020}}     & 2020 & sentence  & tf & Logistic, Multilayer Perceptron, Simple Logistic, SVM, Random Forest, LMT & English   \\ \cline{1-6}

%Nikoloc et al. {\cite{nikolic2020aspect}}     & 2020 & sentence  & BoW, tf*idf    & KNN, SVM, MNB & English   \\ \cline{1-6}% repeated

Mostafa {\cite{Mostafa:2020}}     & 2020 & sentence  & GloVe    & LSTM & English   \\ \cline{1-6}

Nimala \& Jebakumar {\cite{Nimala:2019}}     & 2019 & sentence  & Word Embedding   & LSTM and LSTM with attention layer & English   \\ \cline{1-6}

Srinivas \& Rajendran {\cite{SRINIVAS:2019}}     & 2019 & sentence  & tf*idf   & SVM and kMeans & English   \\ \cline{1-6}

Hariyani et al. {\cite{Hariyani:2019}}     & 2019 & sentence  & tf*idf   & DT, NB, KNN, SVM & English   \\ \cline{1-6}

\end{tabular}

 \caption{A tabular summary of the reviewed studies}
\label{tab:ReviewStudies}
\end{table}

\section{Dataset Resources}
\label{sec:DR}
Most of the researchers in the students' feedbacks SA domains created their own dataset as no benchmark dataset is available publicly. There are two main sources for collecting students' reviews: 
\begin{itemize}
    \item Classrooms' reviews.
    \item E-learning's reviews.
\end{itemize}
In light of online education development, abundant data have produced every day. One of these online educational platforms is Massive Open Online Courses-MOOCs. Many pieces of research conducted on the students' feedback on MOOCs~\cite{Kastrati2020AspectBasedOM,Kastrati2020,Lundqvist2020}.
Marcu et al.~\cite{MARCU2020} for reflecting the characteristics of Romanian high-school students, They collect their data from eleven high schools in Suceava. They asked students age between 16 to 18 answer the question "Please describe in a few words, honestly, how you feel about your school" in a google doc. The answers were in Romanian and for further analysis, they translate them into English. After removing irrelevant feed backs 191 records used for data mining.

Other datasets that used in the students' feedback domain elaborated in the Table~\ref{DatasetstReviewedStudies}.

%%Their assumption was that in some cases if the method tested on the small datasets it does not perform well on %%%the large dataset~\cite{liu2013}

\begin{table}[]
\centering
\begin{tabular}{|p{3.5cm}|p{10cm}|}
\cline{1-2}
References   & Data base  \\ \cline{1-2}
Kastrati et al.{\cite{Kastrati2020}}  & students' reviews that are collected from Coursera    \\ \cline{1-2}
Sangeetha et al. {\cite{Sangeetha2020}} & Vietnamese students feedback corpus (UIT-VSFC)  \\ \cline{1-2}
Nikoli et al. {\cite{Nikoli2020}}    & Official student surveys and online reviews    \\ \cline{1-2}
Sindhu et al. {\cite{Sindhu2019}}    & students feedback of Sukkur IBA University    \\ \cline{1-2}
Kastrati et al. {\cite{Kastrati2020AspectBasedOM}}  & Students Reviews of MOOCs     \\ \cline{1-2}
Shaikh et al. {\cite{Shaikh2019}}    & student feed backs collected at Sukkur IBA University Pakistan  \\ \cline{1-2}
Marcu et al. {\cite{MARCU2020}}     & the opinions of students from eleven high schools in Suceava     \\ \cline{1-2}
Moharil et al. {\cite{Moharil:2020}}     & 6942 students' reviews collected from university feedback form    \\ \cline{1-2}
Estrada et al. {\cite{estrada2020opinion}}     & 2200 words, 24556 opinions, 12084 opinions collected from sentiTEXT, eduSERE, platform SERE, SenttiDict.    \\ \cline{1-2}
Kandhro et al. {\cite{Kandhro:2020}}  & 3000 students' reviews collected from university feedback form   \\ \cline{1-2}
Lwin et al. {\cite{Lwin:2020}} &  3000 students' reviews collected from university feedback form  \\ \cline{1-2}

%Nikolic et al. {\cite{nikolic2020aspect}} &  students' reviews collected from university feedback form  \\ \cline{1-2} repeated
Mostafa {\cite{Mostafa:2020}} &  16175 sentences of students' reviews  \\ \cline{1-2}
Nimala \& Jebakumar {\cite{Nimala:2019}} &  16175 students sentences from survey  \\ \cline{1-2}

Srinivas \& Rajendran {\cite{SRINIVAS:2019}} & Course Forum and Questionnaire \\ \cline{1-2}

Hariyani et al. {\cite{Hariyani:2019}} & Course Questionnaire \\ \cline{1-2}

\end{tabular}
 \caption{Datasets used in the reviewed studies }
\label{DatasetstReviewedStudies}
\end{table}

\section{Research Directions}
\label{sec:rd}
There are several challenges and limitations in students' feedbacks sentiment analysis. Some of the open issues can be listed as follows: 

\begin{itemize}
    \item Lack of benchmark datasets. Despite the diverse nature of datasets, there is no open access dataset and neither any benchmark available. Social networking and collaboration tools used in education  \cite{imran2016analysis} can be valuable source of acquiring students' feedback.
    \item Lack of resources like lexica, corpora, dictionaries for low-resource languages (most of the studies are conducted in English or Chinese language).
    \item Identifying figurative speech like sarcasm, irony, from text.
    \item Generalization - Most of the techniques are domain-specific and thus do not perform well in different domains.
    \item Incapability to handle complex language involving constructs such as double negatives, unknown proper names, abbreviations etc. 
    \item Contextualization and conceptualization of sentiment. Techniques, especially machine learning/deep learning, developed for sentiment analysis need to focus on incorporating the semantic context using sources such as Wordnet, SentiWordnet \cite{weichselbraun2014enriching} or semantically representation using ontologies \cite{kastrati2019impact,Kastrati:2015Sitis} for better grasping one's opinion from the text.
\end{itemize}

\section{Conclusion}
\label{sec:conclusion}
With increasing the demand for online education on one hand and distance study due to the COVID19 pandemic, on the other hand, students' feedback analysis is the most vital task for professors and educational institutes. Hence aspect extraction and sentiments analysis of students' feedback is a hot domain for researchers. There is a huge amount of research conducted on the sentiment analysis of the student feedbacks. Therefore to prevent repetition and giving the latest methods of opinion mining that used in the literature this article focused on the students' feedback papers published in recent years. The findings suggest to create a publicly available benchmark data set, need to incorporate semantics and advanced embedding models for aspect extraction and sentiment analysis.

\bibliography{access.bib}{}
\bibliographystyle{IEEEtran}

\end{document}